\documentclass[english,letterpaper]{smfart}
\usepackage{latexsym,amssymb}%

\usepackage{graphicx}  

\usepackage{epstopdf}

\usepackage{psfrag}

\usepackage{color}

\usepackage{a4}
\usepackage{amssymb,latexsym}
\usepackage[mathscr]{eucal}
\usepackage{cite}
\usepackage{url}

\DeclareFontFamily{OT1}{rsfs}{} \DeclareFontShape{OT1}{rsfs}{m}{n}{ <-7> rsfs5 <7-10> rsfs7 <10-> rsfs10}{}
\DeclareMathAlphabet{\mycal}{OT1}{rsfs}{m}{n}

\newcommand{\mmcK}{\,\,\,\mathring{\!\!\! \mcK}}

\newcommand{\kk}[1]{}

\newcommand{\ohyp}{\,\,\overline{\!\!\hyp}}
\newcommand{\pohyp}{\partial\ohyp}

\newcommand{\regular}{$I^+$--regular}

\newcommand{\eean}{\nonumber\end{eqnarray}}

\newcommand{\mcMext}{\Mext}

\newcommand{\mcHN}{\mcN}

\def\K0{\phi^{K_0}}

\def\X.{\phi^{X}\cdot}

{\catcode `\@=11 \global\let\AddToReset=\@addtoreset}
\AddToReset{equation}{section}
\renewcommand{\theequation}{\thesection.\arabic{equation}}

\newcommand{\fourg}{{\mathfrak g }}

\newcommand{\mcN}{{\mycal N}}

\newcommand{\nopcite}[1]{}

  {\renewcommand{\theenumi}{\roman{enumi}}
   \renewcommand{\labelenumi}{(\theenumi)}}

\newcommand{\mcE}{{\mycal E}}
\newcommand{\mcC}{{\mycal C}}

\newcommand{\nablash}{\nabla{\kern -.75 em
     \raise 1.5 true pt\hbox{{\bf/}}}\kern +.1 em}
\newcommand{\Deltash}{\Delta{\kern -.69 em
     \raise .2 true pt\hbox{{\bf/}}}\kern +.1 em}
\newcommand{\Rslash}{R{\kern -.60 em
     \raise 1.5 true pt\hbox{{\bf/}}}\kern +.1 em}

\newcommand{\mcT}{{\mycal T}}
\newcommand{\mcU}{{\mycal U}}
\newcommand{\mcV}{{\mycal V}}

\newcommand{\hyp}{{\mycal S}}

\newcommand{\mcM}{{\mycal M}}

\newcommand{\mcK}{{\mycal K}}

\newcommand{\bea}{\begin{eqnarray}}
\newcommand{\beaa}{\begin{eqnarray*}}
\newcommand{\bean}{\begin{eqnarray}\nonumber}

\newcommand{\bel}[1]{\begin{equation}\label{#1}}
\newcommand{\beal}[1]{\begin{eqnarray}\label{#1}}
\newcommand{\beadl}[1]{\begin{deqarr}\label{#1}}
\newcommand{\eeadl}[1]{\arrlabel{#1}\end{deqarr}}
\newcommand{\eeal}[1]{\label{#1}\end{eqnarray}}
\newcommand{\eead}[1]{\end{deqarr}}
\newcommand{\eea}{\end{eqnarray}}
\newcommand{\eeaa}{\end{eqnarray*}}

\newcommand{\be}{\begin{equation}}
\newcommand{\ee}{\end{equation}}

\newtheorem{defi}{\sc Coco\rm}[section]

\newtheorem{Theorem}[defi]{\sc Theorem\rm}

\newtheorem{Definition}[defi]{\sc Definition\rm}

\newtheorem{Proposition}[defi]{\sc Proposition\rm}

\def \R {\Reel}

\newcounter{mnotecount}[section]

\renewcommand{\themnotecount}{\thesection.\arabic{mnotecount}}

\newcommand{\mnote}[1]
{\protect{\stepcounter{mnotecount}}$^{\mbox{\footnotesize $%
\!\!\!\!\!\!\,\bullet$\themnotecount}}$ \marginpar{
\raggedright\tiny\em $\!\!\!\!\!\!\,\bullet$\themnotecount: #1} }

\newcommand{\ednote}[1]{}

\definecolor{bluem}{rgb}{0,0,0.5}

\definecolor{mycolor}{cmyk}{0.5,0.1,0.5,0}
\definecolor{michel}{rgb}{0.5,0.9,0.9}

\definecolor{turquoise}{rgb}{0.25,0.8,0.7}
\definecolor{bluem}{rgb}{0,0,0.5}

\definecolor{MDB}{rgb}{0,0.08,0.45}
\definecolor{MyDarkBlue}{rgb}{0,0.08,0.45}

\definecolor{MLM}{cmyk}{0.1,0.8,0,0.1}
\definecolor{MyLightMagenta}{cmyk}{0.1,0.8,0,0.1}

\definecolor{HP}{rgb}{1,0.09,0.58}

\newcommand{\pdoc}{\partial \doc}

\newcommand{\Sext}{\hyp_{\mathrm{ext}}}
\newcommand{\Mext}{\mcM_{\mathrm{ext}}}

\newcommand{\doc}{\langle\langle \mcMext\rangle\rangle}

\def\emph#1{{\it #1}}
\def\textbf#1{{\bf #1}}

\def\AA{{\mycal  A}}

\def\L{{\bf L}}
\def\O{{\bf O}}

\def\R{{\mathbb R}}

\def\K{{\bf K}}

\newcommand{\changedX}{K}

\newcommand{\hahyp}{\,\,\widehat{\!\!\hyp}}%

\def\2{{\overline 2}}

\newcommand{\beqa}{\begin{eqnarray}}
\newcommand{\eeqa}{\end{eqnarray}}

\begin {document}

\frontmatter
\author [P.T.~Chru\'sciel]{Piotr T.~Chru\'sciel\thanks{Supported
 in part by the  Polish Ministry of Science and
Higher Education grant Nr N N201 372736.}}
\address {Vienna University}
 \email{chrusciel@maths.ox.ac.uk}
\urladdr {www.phys.univ-tours.fr/$\sim$piotr}

\author [G.J.~Galloway]{Gregory J.~Galloway\thanks{Supported in part by  NSF grant DMS-0708048}}
\address {University of Miami, Coral Gables}
 \email {galloway@math.miami.edu}
\urladdr {www.math.miami.edu/~galloway}

\title[Uniqueness of static black-holes without analyticity]%
{Uniqueness of static black-holes without analyticity}

\begin {abstract}
We show that the hypothesis of analyticity in
the uniqueness theory of vacuum, or electrovacuum, static black holes
is not needed.
More generally, we show that prehorizons  covering a closed set
cannot occur in well-behaved domains of outer communications.
\end {abstract}
\maketitle


\mainmatter

\section{Introduction}
\label{Sintro}

One of the hypotheses in the current
theory of uniqueness of static vacuum black holes is that of analyticity.
This is used to exclude  null Killing orbits,
equivalently to prove non-existence of \emph{non-embedded degenerate prehorizons covering a closed set},
within the domain of outer-communications; see~\cite{ChCo} for the details.
The aim of this note is to show that analyticity is not needed to exclude such prehorizons,
and therefore can be removed from the set of hypotheses of the classification theorems
in the static case.

More generally, such prehorizons need to, and have been, excluded
in dimension $n+1$ with $n-1$ commuting Killing vectors~\cite{ChHighDim} without assuming analyticity.
Our analysis here provides an alternative, simpler,
approach to this issue for any stationary solutions satisfying the null energy
condition, without the need to invoke more Killing vectors.
(Note, however, that for solutions that are \emph{not static},
all $n-1$ Killing vectors are used to prove that existence of a null
Killing orbit implies existence of a prehorizon.)

In this work we  consider asymptotically flat, or Kaluza-Klein (KK) asymptotically flat
(in the sense of~\cite{CGS}) spacetimes, and show that (for definitions, see below
and~\cite{ChCo}):

\begin{Theorem}
\label{T27II.3}
 $I^+$-regular stationary domains of outer communication $\doc$ satisfying the null
 energy condition do not contain prehorizons, the union of which is closed within $\doc$.
\end{Theorem}

The reader is referred to~\cite{AIK} and references therein for progress towards removing the hypothesis
of analyticity in the general stationary case.

\section{The time of flight  argument}
 \label{S27II10.2}

\begin{figure}[t]
\begin{center} { \psfrag{Mext}{$\phantom{x,}\Mext$}
\psfrag{H}{ } \psfrag{B}{ }
\psfrag{H}{ }
 \psfrag{pSigma}{$\!\!\pohyp\qquad\phantom{xxxxxx}$}
\psfrag{Sigma}{ $\hyp$ }
 \psfrag{toto}{$\!\!\!\!\!\!\!\!\!\!\doc$}
 \psfrag{S}{}
\psfrag{H'}{ } \psfrag{W}{$\mathcal{W}$}
\psfrag{scriplus} {} 
\psfrag{scriminus} {} 
 \psfrag{i0}{}
\psfrag{i-}{ } \psfrag{i+}{}
 \psfrag{E+}{ $\phantom{.}{\mycal E}^+$}
{\includegraphics{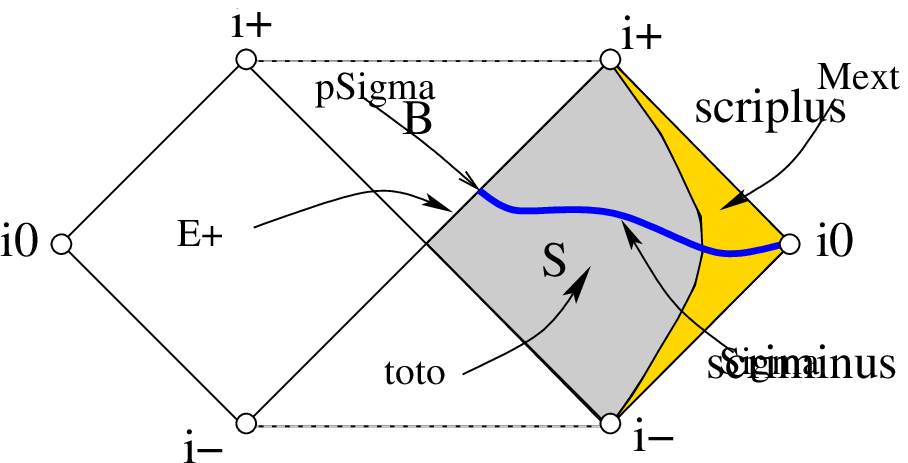}}
}
\caption{The hypersurface $\hyp$ from the definition of \regular ity.
\protect\label{fregu}}
\end{center}
\end{figure}

For the convenience of the reader we recall some definitions from~\cite{ChCo,ChHighDim}:

\begin{Definition}
 \label{Dmain}
Let $(\mcM,\fourg)$ be a space-time containing an asymptotically
flat, or $KK$--asymptotically flat end $\Sext$,
and let  $\changedX $ be a stationary Killing vector field  on $\mcM$.
We will say that $(\mcM,\fourg,\changedX)$ is $\mbox{\rm {\regular}}$%
\index{$\mbox{\rm {\regular}}$}
if $\changedX $ is complete, if the domain of outer communications
$\doc$ is globally hyperbolic, and if $\doc$ contains a spacelike,
connected, acausal hypersurface $\hyp\supset\Sext $,%
\index{$\hyp$}
the  closure $\ohyp $ of which is a topological manifold with boundary,
consisting of  the union of a compact set and of a finite number of
asymptotic ends, such that the boundary $ \pohyp:= \ohyp \setminus \hyp$
is a  topological manifold satisfying
\bel{subs}
\pohyp \subset \mcE^+:= \partial \doc \cap I^+(\Mext)
 \;,
 \ee
with $\pohyp$
meeting every generator of $\mcE^+$ precisely once. (See Figure~\ref{fregu}.)
\end{Definition}

The definition appears to capture the essential ingredients required for a successful
classification of vacuum~\cite{ChCo} or electro-vacuum~\cite{Costaelvac}
black holes. Whether or not the definition is optimal
from this point of view remains to be seen.
In any case, one of its consequences is the
\emph{structure theorem}~\cite{ChCo,ChHighDim}, which in essence
goes
back to~\cite[Lemma~2]{ChWald}, and
which represents $\doc$ globally as $\R\times\hyp$,
with the Killing vector tangent to the $\R$ factor.

Another notion that is essential for the current work is:

\begin{Definition}%
\index{Killing prehorizon}%
 \label{DefKH}
Let $\changedX $ be a Killing vector and set
\bel{KHd}
 \mcHN[\changedX ]:=\{ p\in \mcM \ |  \ \fourg(\changedX ,\changedX )|_p= 0\;,\ \changedX|_p  \ne 0\}\;.
\ee
Every connected, not necessarily embedded,  null hypersurface
$\mcN_0\subset\mcHN[\changedX ]$ {to which $\changedX $ is tangent}
will be called a  $\mbox{\rm Killing prehorizon}$.
\end{Definition}

It follows from~\cite[Corollary~3.3 and Lemma~5.14]{ChCo} that in vacuum \regular\
space-times which are static,
or four-dimensional stationary and axisymmetric, or $(n+1)$--dimensional with $n-1$ commuting Killing
vectors, the set covered by Killing prehorizons associated with a Killing vector field $\changedX$ is closed
within $\doc$. This remains true for electrovacuum space-times in dimension $3+1$.

For further purposes it is convenient to introduce

\begin{Definition}
Let $(\mcM,g)$ be a space-time with a complete Killing vector field $\changedX$,
and let $\Omega\subset \mcM$. We shall say that a closed set $\mcK\subset \Omega$
is an {\rm invariant quasi-horizon in $\Omega$} if $\mcK$ is a union of
pairwise disjoint null (not necessarily embedded) hypersurfaces,  called leaves.
We further assume that the leaves of $\mcK$ are invariant under the flow of $\changedX$,
and that every null geodesic maximally extended within $\Omega$ and
initially tangent to a leaf of $\mcK$ remains on $\mcK$.
\end{Definition}

From what has been said, it follows that

\begin{Proposition}
 In static, or stationary axi-symmetric, $I^+$-regular vacuum space-times,
the union of Killing prehorizons forms a (possibly empty)
spatially bounded quasi-horizon in $\doc$.
\end{Proposition}

Here ``spatially bounded" means that it does not extend infinitely far
out in the end $\{0\} \times \Sext$. In cases of interest the Killing
vector flow acts as a translation along the $\R$ factor, so in fact one has a $t$-independent
bound on the extent on each slice $\{t\} \times \Sext$.

Theorem~\ref{T27II.3} follows now from:

\begin{Theorem}
 \label{Twhich}
Consider an asymptotically flat, or $KK$--asymptotically flat,
globally hyperbolic domain of outer communications $\doc$, satisfying the null energy condition,
diffeomorphic to $\R \times \hyp$,
with the Killing vector, tangent to the $\R$ factor, approaching a
time translation in the asymptotic
region.
Then there are no invariant quasi-horizons in $\doc$.
\end{Theorem}

\noindent{\sc Proof of Theorem~\ref{Twhich}:}
Let $R\in \R$ be large enough so that the constant-time spheres lying on the
timelike hypersurface
$$
 \mcT:= \R\times \{|\vec x| =R\}
$$
are both \emph{past} and \emph{future inner trapped}, as defined in~\cite{CGS}.
Without loss of generality we can assume that $\mcK$ does not intersect the region
$\{|\vec x| \ge R\}$; indeed,  $\changedX$ cannot be tangent
to the null leaves of $\mcK$ in the asymptotically flat region, where it is timelike.
Let $\mcC$ denote the following class of causal curves:
\beaa
 \mcC&:=&\{ \gamma \ | \ \mbox{$\gamma:[0,1]\to \mcM$
 is a causal curve which starts and ends}
\\
 &&
 \mbox{  at $\mcT$, and meets
 $\mcK_0:=\mcK\cap (\{0\}\times \hyp)$ }\}
 \;.
\eeaa
The \emph{time of flight } $\tau_\gamma$ of $\gamma$ is defined as
$$
 \tau_\gamma:= t(\gamma(1))- t(\gamma(0))
 \;,
$$
where $t$ is the time-function associated with the decomposition $\mcM= \R \times \hyp$.
We write $\hyp_\tau$ for $t^{-1}(\tau)\equiv \{\tau\}\times \hyp$.

Let $\mathring \tau$ denote the infimum of $\tau_\gamma$ over $\gamma\in \mcC$.
We wish to show that if $\mcK_0$ is non-empty, then
$\mathring \tau$ is attained on a smooth null geodesic $\mathring \gamma$, with a) initial and end
points on $\mcT$, b) meeting $\hyp_0$ at $\mcK_0$, c) meeting $\mcT$ normally to the level sets
of $t$.

In order to construct $\mathring \gamma$,
let $\gamma_i\in\mcC$ be any sequence of causal curves
such that $\tau_{\gamma_i}\to \mathring
\tau$. Let   $\gamma$ be any
causal curve in $\mcC$, then $0>t(\gamma_i(0))\ge -\tau_\gamma $ and
$0<t(\gamma_i(1))\le \tau_\gamma$ for $i$ large enough.
Hence for $i$ large enough all the $\gamma_i(0)$'s belong to
the compact set $[-\tau_\gamma,0]\times \{|\vec x|=R\}$;
similarly the $\gamma_i(1)$'s belong to
the compact set $[0,\tau_\gamma]\times \{|\vec x|=R\}$.
By global hyperbolicity there exists an accumulation curve $\mathring \gamma$
of the $\gamma_i$'s which is a $C^0$ causal curve.

Since $ \mcK_0$ is closed in $\doc$,
$\mathring\gamma$ meets $\mcK_0$ at some point $\mathring p$.
It is standard that $\mathring \gamma \cap \{t<0\}$ is
a smooth null geodesic, since otherwise $\mathring p$ would be timelike related to
$\mathring \gamma (0)$, which would imply existence of a  curve in $\mcC$
with time of flight less than $\mathring \tau$. Similarly,  $\mathring \gamma \cap \{t>0\}$ is a smooth
null geodesic.

Next, in a similar fashion (see~\cite[Lemma~50, p.~298]{BONeill}),
$\mathring \gamma$ meets $\mcT_{t(\mathring \gamma (0))}$
and  $\mcT_{t(\mathring \gamma (1))}$ orthogonally, where
$
 \mcT_\tau:= \hyp_\tau \cap \mcT
  \;.
$

We claim that $\mathring \gamma$ is
also smooth at $\mathring p$. To see that, let $\mmcK $ denote that leaf of $\mcK$
that passes through $\mathring p$. Then the portion of $\mathring \gamma$ that lies to the
causal past of $\mathring p$
must meet $\mmcK $ transversally: Otherwise $\mathring\gamma\cap J^-(\mathring p)$
would coincide with that portion of the null Killing orbit of $\changedX$ through $\mathring p$
that lies to the past of $\mathring p$, but those never
reach $\mcT$, since $\mcK$ is spatially bounded.
Similarly
the portion of $\mathring \gamma$ that lies to the
causal future of $\mathring p$
must meet $\mmcK $ transversally. Suppose that the two geodesic segments
forming $\mathring \gamma$
do not join smoothly at $p$. Then there exist arbitrary small deformations of $\mathring \gamma$
which produce a timelike curve with the same end points as $\mathring \gamma$, and
hence the same time of flight. By transversality, and because
there exists a small neighbourhood $\mcV$ of $\mathring p$ in which
the connected component of  $\mmcK\cap \mcV $
passing through $\mathring p$
forms a null embedded hypersurface,
any such deformation, say
$\hat \gamma$, will
meet $\mmcK $ at some point $\hat p$. Let $\phi_t$ denote the flow of $\changedX$,
then
$$
 \bar \gamma:=\phi_{-t(\hat p)}(\hat \gamma)
 $$
is a timelike curve in $\mcC$ which has the same
time of flight as $\mathring \gamma$. Since $\bar \gamma$ is timelike, it can be
deformed to a causal curve with shorter time of flight.
This contradicts the definition
of $\mathring \tau$, and hence proves a), b) and c).

Let $\tau_*=t(\mathring \gamma(0)) $. We claim that d) $\mathring \gamma  $
minimizes the time of flight
amongst \emph{all} nearby differentiable causal curves from $\mcT_{\tau_*}$ to $\mcT$.
Indeed, by transversality of $\mathring \gamma$ to $\mmcK $,
there exists a neighbourhood $\mcU$ of $\mathring \gamma$ in the space
of differentiable curves such that every curve $\gamma$
in this neighbourhood intersects $\mmcK $.
Suppose, then,  that there exists a causal curve $\gamma \in \mcU$
which starts at $\mcT_{\tau_*}$, ends at $\mcT$,
and has time of flight smaller  than $\mathring \tau$. Then $\gamma$ intersects $\mmcK $ at
some $p$. But then $\phi_{-t(p)}(\gamma)$
is in $\mcC$ and has time of flight smaller than $\mathring \tau$,
which contradicts the definition of $\mathring \tau$, whence d) holds.

This provides a contradiction to $\mcK$ being non-empty, as there are no
causal curves with the property d) by~\cite[Proposition~3.3]{CGS}.
\qed

\section{Non-rotating horizons and maximal hypersurfaces}
 \label{S27II10.1}

In this section we provide an alternative simple argument
to exclude prehorizons within the domain of outer communication,
which applies to four-dimensional static vacuum space-times.

Let $(\mcM,\fourg)$ be an asymptotically flat, $I^+$-regular, vacuum
space-time with a \emph{hypersurface orthogonal} Killing vector $\changedX$.
By~\cite{CRT} all components of the future event horizon $\mcE^+$ are non-degenerate. We can
therefore carry-out the construction of~\cite{RaczWald2} if necessary to obtain that $\pdoc$
is the union of
bifurcate Killing horizons. By~\cite{ChWald1}, $\doc$ contains a maximal Cauchy hypersurface $\hyp$.
By~\cite{Sudarsky:wald} (compare the argument at the end of \cite[Section~7.2]{ChCo}), $\hyp$ is totally geodesic.
Decomposing $\changedX$ as $\changedX= Nn +Y$, where $n$ is the field of future-directed unit normals to $\hyp$,
and where $Y$ is tangent to $\hyp$, one finds from the Killing vector equations that
$$
 D_i Y_j + D_j Y_i = -2 NK_{ij}
 \;.
$$
But the right-hand-side vanishes,
thus $Y$ is a Killing vector of the metric $\gamma$ induced on $\hyp$ by $g$. Now, $Y$ is
asymptotic to zero as one recedes to infinity in $\Mext$,
hence $Y=0$ by usual arguments, (see, e.g., the proof
of~\cite[Proposition~2.1]{ChMaerten}). Since $\changedX$ has no zeros within $\doc$ we conclude
that $N$ has no zeros on $\hyp$.
Alternatively, $N$ satisfies the equation
\bel{Neq1}
\Delta N = K^{ij}K_{ij} N\;,
\ee
vanishes on $\partial \hyp$, and is asymptotic to one as one recedes to infinity along
the asymptotically flat region,
and thus has no zeros by the strong maximum principle.
Whatever the argument, $\changedX=Nn$ is timelike everywhere on $\doc$, and there
are no prehorizons within $\doc$.

The above argument applies \emph{verbatim} to higher dimensional vacuum metrics, as well as to
four-dimensional electrovacuum metrics, for
configurations where all horizons are
non-degenerate. A proof of existence of maximal hypersurfaces with sufficiently controlled
asymptotic behaviour near the degenerate horizons would extend this argument to
the general case. In any case, the proof based on the time of flight covers more general situations.

\section{Conclusions}
 \label{S27II10.3}

Recall that  a manifold $\hahyp$ is said to be of \emph{positive
energy type}
\index{positive energy type}%
if there are no asymptotically flat complete Riemannian
metrics on $\hahyp$ with nonnegative scalar curvature and vanishing mass
except perhaps for a flat one. This property has been proved so far for all
$n$--dimensional manifolds $\hahyp$ obtained by removing a finite
number of points from a compact manifold of dimension  $3\le n\le
7$~\cite{SchoenCatini}, or under the hypothesis that $\hahyp$ is a spin
manifold of any dimension $n\ge 3$, and is expected to be true in
general~\cite{ChristLohkamp,Lohkamp}.

Using results already established elsewhere~\cite{ChCo,CostaPhD,Costaelvac,CT,Chstatic}
together with Theorem~\ref{T27II.3} one has:

\begin{Theorem}
 \label{T27II.1}
Let $(\mcM,\fourg)$ be a  vacuum $n+1$ dimensional space-time, $n\ge 3$,
containing a spacelike, connected, acausal
hypersurface $\hyp $, such that $\ohyp $ is a topological manifold with
boundary, consisting of  the union of a compact set and of a finite number
of asymptotically flat ends. Suppose that there exists on $\mcM$ a
complete static Killing vector $\changedX$, that $\doc$ is globally
hyperbolic, and that $ \pohyp\subset\mcM\setminus \doc$.
Let  $\,\,\widehat{\!\! \hyp}$ denote the manifold obtained by
doubling $\hyp$ across the non-degenerate components of its boundary
and compactifying,  in the doubled manifold, all  asymptotically flat regions
but one to a point. If $\,\,\widehat{\!\! \hyp}$ is of positive energy type, then   $\doc$
is isometric to the domain of outer communications of a Schwarzschild space-time.
\end{Theorem}

\begin{Theorem}
 \label{T27II.2}
Under the remaining hypotheses of Theorem~\ref{T27II.1} with $n=3$, suppose instead that
$(\mcM,\fourg)$ is electrovacuum with the Maxwell
field invariant under the flow of $\changedX$. Then  $\doc$
is isometric to the domain of outer communications of a Reissner-Nordstr\"om
or a standard Majumdar-Papapetrou space-time.
\end{Theorem}

\noindent {\sc Acknowledgements:}
PTC is grateful to the University of Miami for hospitality and support during part
of work on this paper.

\bibliographystyle{amsplain}
\bibliography
{
../references/newbiblio,%
../references/newbib,%
../references/reffile,%
../references/bibl,%
../references/Energy,%
../references/hip_bib,%
../references/netbiblio,%
../references/addon}
\end {document}